\documentclass[preprint,11pt]{JHEP3} 

\JHEPspecialurl{http://jhep.sissa.it/JOURNAL/JHEP3.tar.gz}

\usepackage{epsfig,multicol,amsmath}

\usepackage{epstopdf}
\usepackage{rotating}
\usepackage{cite}

\def\beq{\begin{equation}}
\def\eeq{\end{equation}}
\def\bea{\begin{eqnarray}}
\def\eea{\end{eqnarray}}

\def\eq#1{{Eq.~(\ref{#1})}}
\def\fig#1{{Fig.~\ref{#1}}}

\newcommand{\as}{\alpha_S}

 \newcommand{\SP}{\langle \mid S^2 \mid\rangle}

\newcommand{\Lb}{\left(}
\newcommand{\Rb}{\right)}
\parskip=\itemsep               

\newcommand{\y}{{\cal Y}}

\def\pom{{I\!\!P}}
\def\reg{{I\!\!R}}
\title{Soft interactions at high energies: QCD motivated approach}
\author{\Large  E. Gotsman\thanks{Email:
gotsman@post.tau.ac.il.}\,, E. Levin\thanks{Email:
leving@post.tau.ac.il,
levin@mail.desy.de.}\,, U. Maor\thanks{Email: maor@post.tau.ac.il.}\, 
\,and\, J.S. Miller\thanks{Email: jeremymi@post.tau.ac.il.}\\
Department of Particle Physics, School of Physics and Astronomy\\
Raymond and Beverly Sackler Faculty of Exact Science\\  
Tel Aviv University, Tel Aviv, 69978, Israel}

\abstract{We propose a QCD motivated theoretical approach
 to  high energy soft interactions, which successfully describes the
 experimental data on total, elastic and diffraction cross sections. 
 We predict that  the survival probability for the diffractive
 Higgs production at the LHC energy is small (less than 1\%),
 and investigate the influence of suggested  corrections e.g. threshhold 
effects and semi-enhanced diagrams,
 on this  value.}
\dedicated{PACS: 13.85.-t, 13.85.Hd, 11.55.-m, 11.55.Bq}
\preprint{TAUP -2878-08\\
{\tt 0805.2799 [hep-ph]}\\
\today}
\begin{document}
\section{Introduction}
For the past three decades the physics of soft  
interactions has been considered an unsavory topic. Indeed, the relation 
between  high energy phenomenology and the microscopic theory: QCD, has been
neglected, as there has been no progress in solving the problem of the 
confinement of quarks and gluons. 
The goal of this letter is to present an attempt  to construct a 
self consistent theoretical approach
to strong interactions at high energy, which provides a natural bridge to 
the perturbative QCD description at short distances. 
\FIGURE[h]{\begin{minipage}{90mm}
{\centerline{\epsfig{file=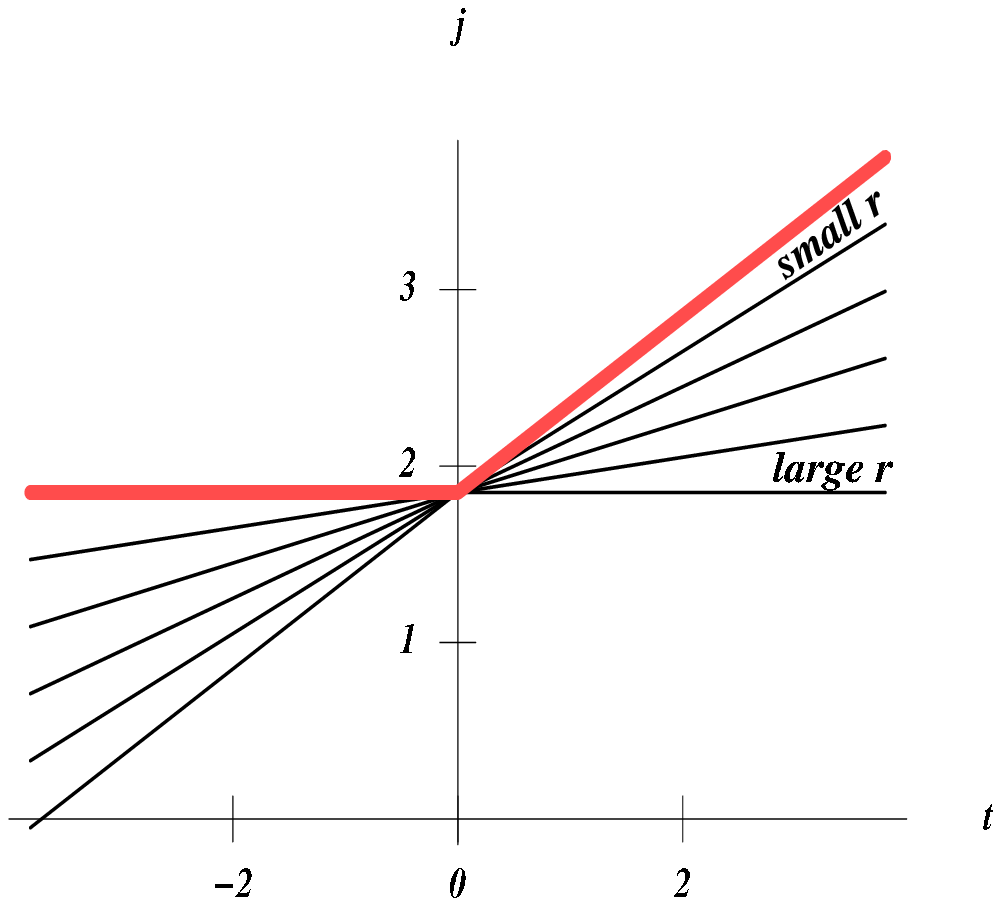,width=85mm}}
\caption{The behaviour of the Pomeron trajectory in N=4 SYM accordingly 
to Ref.\cite{BST}. The figure is taken from Ref.\cite{BST}}
\label{pom4}}
\end{minipage}}

Following on the results of our paper\cite{GLMM}, we question the  
widely held
prejudice that high energy scattering stems from  long distances.
We believe, that the only dimensional scale that governs
the high energy interaction, is the saturation momentum ($Q_s$),
which appears both in pQCD
at high parton density\cite{GLR,MUQI,B,K}, and in the non-perturbative 
approach to QCD  based on the Colour Glass Condensate\cite{MV,JIMWLK}).
The saturation momentum increases at high energies and, 
therefore, we believe
that the higher  the energy, the shorter are the distances
involved in the interaction.
The key argument supporting
our hypothesis is the fact that in fits to data,  the slope of
the Pomeron trajectory ($\alpha'_\pom$)  
is very small\cite{GLMM,KMRNEW,KMR5}. 
At first sight a small   $\alpha'_\pom$
contradicts one of the principle results of Regge theory, i.e.
that all resonances lie on Regge trajectories.
However, the first theoretical analysis of the  Pomeron's structure
in a theory in which we can treat
the region of strong couplings
(the strongly coupled N=4 super Yang-Mills theory
with AdS/CFT correspondence),
given in Ref.\cite{BST}, shows that
$\alpha^{'}_{\pom} \to 0$ for $t<0$.
This result is compatible with our  approach.
In the same theory $\alpha^{'}_{\pom}  >  0 $  for $t>0$,
which is in reasonable agreement with Regge theory and high energy
phenomenology (see \fig{pom4} which was taken from Ref.\cite{BST}).
Therefore, the small value of $\alpha^{'}_{\pom}$ 
in the scattering region is not only possible, but is  
a requisite, so as to be consistent 
with a strong coupling theory.

In\,\, perturbative\,\,QCD\,\,$\alpha^{'}_{\pom}$\,\, for
a\,\, BFKL \,\,\,\,Pomeron\cite{BFKL}, 
which determines the high energy scattering amplitude,
is equal to zero, since 
$\alpha^{'}_{\pom} \,\,\propto\,1/Q^2_s \xrightarrow{s \to \infty} 0$. 
$W = \sqrt{s}$ is the energy in the c.m. frame.
Therefore, the assumption that 
$\alpha^{'}_{\pom} \to 0 $, provides the
necessary condition so as to link the strong interactions 
with the hard interactions that are described by pQCD.

The only microscopic explanation of Regge theory is provided by 
the partonic approach\cite{GRIB,FEYN}, in which 
the slope of the Pomeron trajectory is related to the mean transverse
momentum of the exchanged partons, 
$\alpha'_\pom\,\propto\,\,1/<p_t>$.  
$<p_t>$ is the mean parton momentum.
The commonly held view from  high energy phenomenology, is
that $\alpha'_\pom \,=\,0.25 \,GeV^{-2}$\cite{DL}. 
 
The phenomenological value of $\alpha'_\pom$ 
stems from the supposition, that
the shadowing corrections are small, and the shrinkage of 
the forward elastic peak is induced by the slope of the Pomeron trajectory.
The above supposition is not compatible with the high value of 
$\alpha_{\pom}(0)$ implied by the ISR-Tevatron cross section data.
In Ref.\cite{GLMM} an alternative option was suggested in which  
eikonal shadowing corrections are significant 
yielding a diminishing $\alpha'_\pom  = 0.01 \, GeV^{-2}$ value.  
This result is not affected by a more complicated form 
of the Pomeron interactions. 
A similar result was obtained, also, in Ref.\cite{KMRNEW} 
in which the experimental elastic 
data was reproduced with $\alpha'_\pom  = 0$  
\footnote{In their last five papers (see Refs.\cite{KMR5})
the Durham group has considered a diminishing, non zero, 
$\alpha'_\pom$.}.
We conclude that a small $\alpha'_\pom$
is a reasonable, data compatible, option which induces a large 
typical parton momentum. 
The resulting running QCD coupling 
$\as = \pi/b \ln\Lb <p^2_t>/\Lambda^2_{QCD}
\Rb \,\ll\,1$. We shall consider it as
the small parameter in the pQCD estimates of 
the Pomeron-Pomeron interaction vertices.
We wish to stress that, regardless of its intuitive appeal, 
the parton model does not provide a 
reliable approach,{\bf  since it presumes a short range 
interaction in rapidity, while the exchange of gluons in QCD generate 
 long range rapidity interactions}. Nevertheless, the model 
convincingly illustrates, that the smallness of $\alpha'_{\pom}$, 
implies that the typical interaction distances are short. 
As we have discussed, we estimate that they are of the order of $1/Q_s$.

The second  assumption underlying our approach is that only the 
triple Pomeron interaction is essential at high energy. This does 
not precludes multi Pomeron vertices which are estimated by fan diagrams. 
As such these vertices are not local in rapidity.
This  has been proven in 
pQCD (see Refs.\cite{BART,BRN}).
This selection rule (considering only the triple Pomeron vertex) 
postulates that in the leading log $x$ approximation of 
pQCD, with a large number of colours ($N_c \,\gg\,1$), all
other possible interactions are suppressed. 
It is well known that at high energy, and
for $N_c \gg 1$, the correct degrees of freedom are
the colourless dipoles \cite{MUCD}. In terms of these
dipoles, the selection rule reads, that only the decay
of one dipole to two dipoles contributes to the high energy amplitude.
This implies that a colourless dipole
does not interact with the target via multi Glauber-type 
rescatterings, but only through the splitting of one dipole  into
two dipoles and the  merging of two dipoles to one dipole.    
This postulate provides a natural bridge
between our theoretical approach to high energy scattering, 
and the corresponding pQCD high energy scattering amplitude.

In this paper we present a theoretical
approach for soft interactions at high
energies, incorporating the following assumptions:
\begin{enumerate}
\item \quad $\alpha'_\pom  = 0$.
\item \quad The Pomeron interactions reduce via fan diagrams to 
triple Pomeron vertices.
\item \quad $\alpha'_\reg \neq 0$ and it's slope,
as well as other parameters of the Reggeons, should be determined from 
experiment. 
\item \quad A Pomeron - hadron interaction is treated 
phenomenologically using the eikonal formulae, 
in the framework of a two channel formalism, based on the
Good-Walker mechanism \cite{GW}.
\end{enumerate}

The first two assumptions allow us to build a theory of interacting 
Pomerons, while the last two points are phenomenological.
\FIGURE[ht]{
\centerline{\epsfig{file=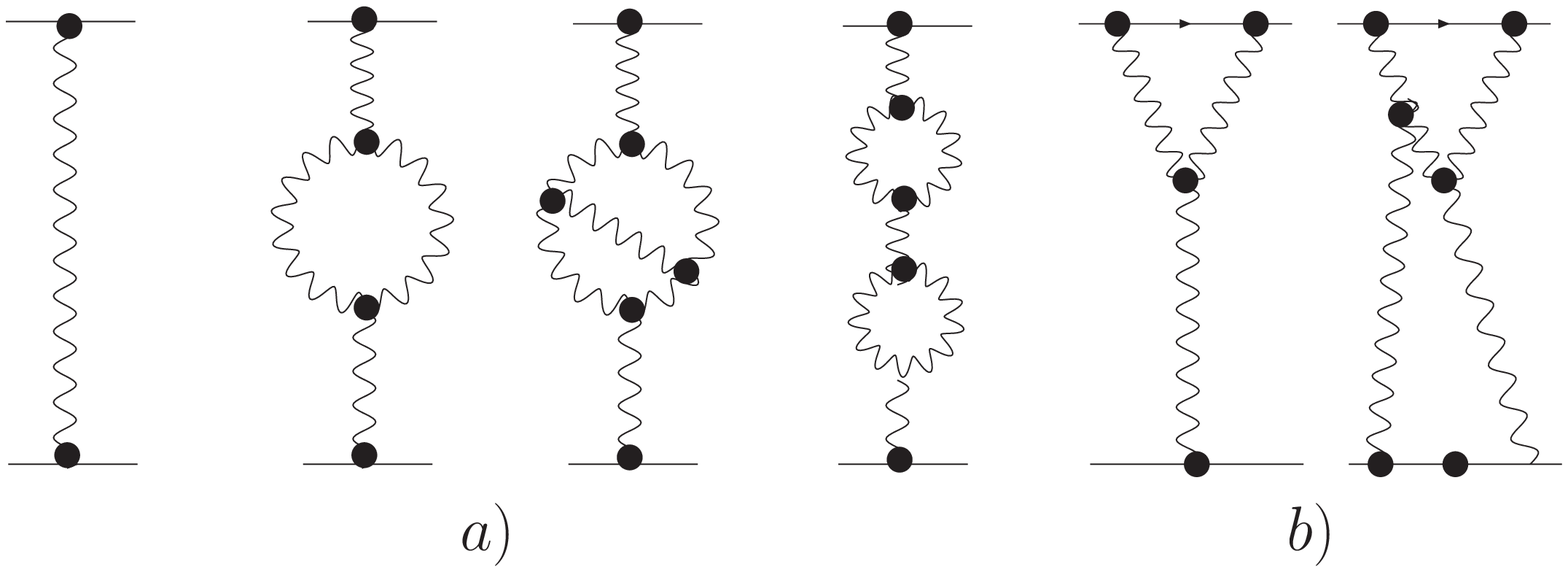,width=165mm,height=50mm}}
\caption{ The different contributions to  Pomeron Green's function ( 
enhanced diagrams of \fig{enh}-a) and the 
semi-enhanced diagrams that has not been taken into account in the 
Pomeron Green's function.} 
\label{enh}}
Details of the two channel (Good-Walker mechanism, 
see Ref.\cite{GW}) that we use for 
describing the elastic and diffractive scattering eigenstates are given in 
Ref.\cite{GLMM}.
\section{The Pomeron interaction (Pomeron Green's function)}
Our key idea is to use the general results of pQCD
relating to high energy scattering\cite{QCDHE}. Consequently:\\
1) In the leading order approximation of pQCD, only one
Pomeron splitting into two Pomerons, and two Pomerons merging into one
Pomeron are taken into account\cite{BART,BRN}. All other Pomeron
vertices do not explicitly appear
in the leading log $x$ approximation of pQCD.
We, therefore, restrict ourselves to  
sum only Pomeron diagrams containing triple Pomeron vertices.\\
2) We consider a theory with $\alpha^{\prime}_{\pom}\,\,=\,\,0$.\\
3) Using our approach, we can predict the anticipated values of all 
Pomeron interactions: 
the intercept of the Pomeron above unity $\Delta_\pom$ is 
$\,\propto \,\as$, and the triple Pomeron vertex
coupling $g_{3\pom}\,\propto\,\as^2$.

The theory which includes all the above ingredients can
be formulated in terms of a generating function\cite{MUCD,LELU}
\beq \label{Z}
Z(y,\,u)\,\,=\,\,\sum_n\,\,P_n(y)\,\,u^n, 
\eeq
where, $P_n(y)$ is the probability to find $n$-Pomerons (dipoles) at 
rapidity $y$.
At rapidity $y\,=\,Y=\ln(s/s_0)$  we can impose
an arbitrary initial condition. For example, demanding that 
there is only one fastest parton (dipole), which is $P_1(y\,=\,Y)\,=\,1$,
while $P_{n>1}(y\,=\,Y)\,=\,0$. In this case we have the following
initial condition for the generating function
\beq \label{INC1}
Z(y\,=\,Y)\,=\,u\,.
\eeq
At $u =1$ 
\beq \label{INC2}
Z(y,\,u\,=\,1)\,\,=\,\,1,
\eeq
which follows from the physical meaning of $P_n$ as a probability.
The solution, with these two conditions, will give us the sum of enhanced 
diagrams (see \fig{enh}-a).

For the function $Z\Lb u\Rb $ the following simple equation can be written 
(see Ref.\cite{GLMM} and references therein)
\beq \label{GFEQ}
\,\,-\frac{\partial\,Z(y,\,u)}{\partial\, \y}\,\,
=\,\,-\,\Gamma(1 \to 2)\,u\,(1\,-\,u) 
\,\,\frac{\partial\,Z(y,\,u)}{\partial\, u}\,\,\,+\,\,\,
\Gamma(2 \to 1)\,u\,(1\,-\,u) \,\,
\frac{\partial^2\,Z(y,\,u)}{\partial^2\,u},
\eeq
where, $\Gamma(1 \to 2)$ describes the decay of one Pomeron (dipole)
into two Pomerons (dipoles), while 
$\Gamma(2 \to 1)$ relates to the merging of
two Pomerons (dipoles) into one Pomeron (dipole).

The first term on the r.h.s. describes two processes:
the term proportional to $u$
can be viewed as a probability of a dipole annihilation
in the rapidity range $( y $ to $ y - dy )$ (death term).
The term with factor $u^2$ is a
probability to create one extra dipole (birth term). 
The second term on the r.h.s.  describes the same type of
processes for Pomeron (dipole) merging.

The description of the parton system given by \eq{GFEQ}, is equivalent to
the path integral formulation of the Pomeron interaction
(see Ref.\cite{GLMM}). The path integral formulation of the 
Pomeron interaction leads to the automatic inclusion  
of  $t$-channel unitarity constraints, while
in the generating function formulation, 
$s$-channel unitarity has been taken into account.

Using the functional $Z$,
we  find the scattering amplitude\cite{K,LELU}, using the following formula:
\beq \label{AM}
N\Lb Y\Rb\,\,\,\equiv\,\,\,\mbox{Im} A_{el}\Lb Y\Rb \,\,\,
=\,\,\,\sum^{\infty}_
{n =1}\frac{(-1)^n}{n!}\,\,\,
\frac{\partial^n\,Z(y,\,u)}{\partial^n\, u}|_{u =1}
\,\gamma_n(Y=Y_0,b),
\eeq
where, $\gamma_n(Y=Y_0,b)$ is the scattering amplitude of 
$n$-partons (dipoles) at low energy. 
These amplitudes depend on the impact 
parameters which are the 
same for all $n$ partons since
$\alpha'_\pom\,\,=\,\,0$. Consequently, we
neglect the diffusion of partons in impact parameter space.
\eq{AM} corresponds to the partonic approach\cite{GRIB,FEYN}, in which
a high energy scattering process can be viewed as a two stage sequence. 
The first stage is the  development of the partonic wave function, 
which we consider by introducing the generating
function $Z$. The second stage is given by \eq{AM} corresponds to  
the interaction of the lowest energy partons
('wee' partons) with the target, which is described by the amplitudes 
$\gamma_n(Y=Y_0,b)$. Assuming that there are no correlations between the
interacting partons (dipoles) at low energy, we take
$\gamma_n(Y=Y_0,b)\,\,=\,\,\gamma^n_1(Y=Y_0,b)$\cite{K}.

The generating function approach given by \eq{Z}, \eq{GFEQ} and \eq{AM},
has the advantage that it can be solved analytically 
(see Ref.\cite{KOLE}). This 
solution leads to a constant cross section at high energy, while the
interaction without the four Pomeron term, decreases at 
high energy\cite{AMCP}(see also Refs. \cite{GRPO,BMMSX}). 
The problem is  that the inclusion of the four Pomeron term  
contradicts our QCD motivated approach. 
This highlights the importance of $s$-channel unitarity studies of 
the asymptotic high energy behaviour of the scattering amplitude. 
The above problem is not relevant to our present investigation as  
we restrict the energy range where we can trust our approach, 
to energies at which the four Pomeron term is not yet significant. 
As discussed in our paper\cite{GLMM} this range is $W \leq 10^5GeV$.

We wish to emphasis  that the form of the equation, 
as well as the functional formulation of the Pomeron interaction, 
is the same as that for interacting dipoles in QCD. 
The only difference is that in QCD the size of the dipole
can be changed due to an interaction, while in our case the size
of the dipole is fixed.
The idea that the transition from short distances
to long distances goes through the stage of freezing
the typical size of the interacting dipoles, has been 
around for a long time, and has both theoretical
and phenomenological justification (see, for example, 
Refs.\cite{KHLE,KHLEK,KAORSAY,KOP}).  
This gives us further confidence that our approach provides
a natural link with the perturbative QCD approach.

Unfortunately, the
exact solution to \eq{GFEQ} with arbitrary initial and 
boundary conditions, is not very transparent.
In a number of papers \cite{MPSI,LE,KOLE1,LEPR,LMP,GLMM},
it has been shown  that, using Mueller-Patel-Salam-Iancu
approximation, one can simplify the solution, and  reduce it
to a simple analytical form.
This approximation yields a solution with an accuracy of the order
of $\gamma\,\exp\Lb- \Delta_\pom\,Y\Rb\,\,
\propto \as^2\,\exp\Lb-\as Y\Rb\,\lll \,1$.

In ref.\cite{GLMM}, and in this letter, we only calculate 
the Green's function of the Pomeron, neglecting for
example,  diagrams of the \fig{enh}-b type.
The reason for this procedure is that the interaction of the
Pomeron described by the exact Green's function, will be simpler
than the interaction of the bare Pomerons
(see Ref.\cite{GRIBRT} in which this approach was suggested).

The exact expression for the Pomeron Green's function has the form
\beq \label{AMMPSI}
G_\pom \left(Y\right)\,\,\,=\,\,\,1\,\,-\,\,\exp \Lb \frac{1}{T(Y)}\Rb\,
\frac{1}{T(Y)}\,\,
\Gamma\Lb 0,\frac{1}{T(Y)} \Rb,
\eeq
where $\Gamma \Lb 0,x \Rb$ is the incomplete gamma function 
(see {\bf 8.350 - 8.359} in Ref.\cite{RY}) and
\beq \label{T}
T\Lb Y\Rb\,\,\,=\,\,\gamma\,e^{\Delta_{\pom}\,Y}.
\eeq
where, $\gamma$ has a simple meaning of being the amplitude of 
of the two dipoles interaction at low energy.
The derivation of \eq{AMMPSI} as well as derivation of the formulae for 
the cross sections of the single and double diffraction production,
are given in Ref.\cite{GLMM}.
\section{The results of the fit}
The pertinent details of our fit to the experimental data, and our 
determination of the relevant parameters of the model, needed to describe 
the soft interactions, are contained in \cite{GLMM}. In this section we 
will only mention the salient features, and results of the fit. 
\FIGURE[h]{\begin{minipage}{90mm}
{\centerline{\epsfig{file=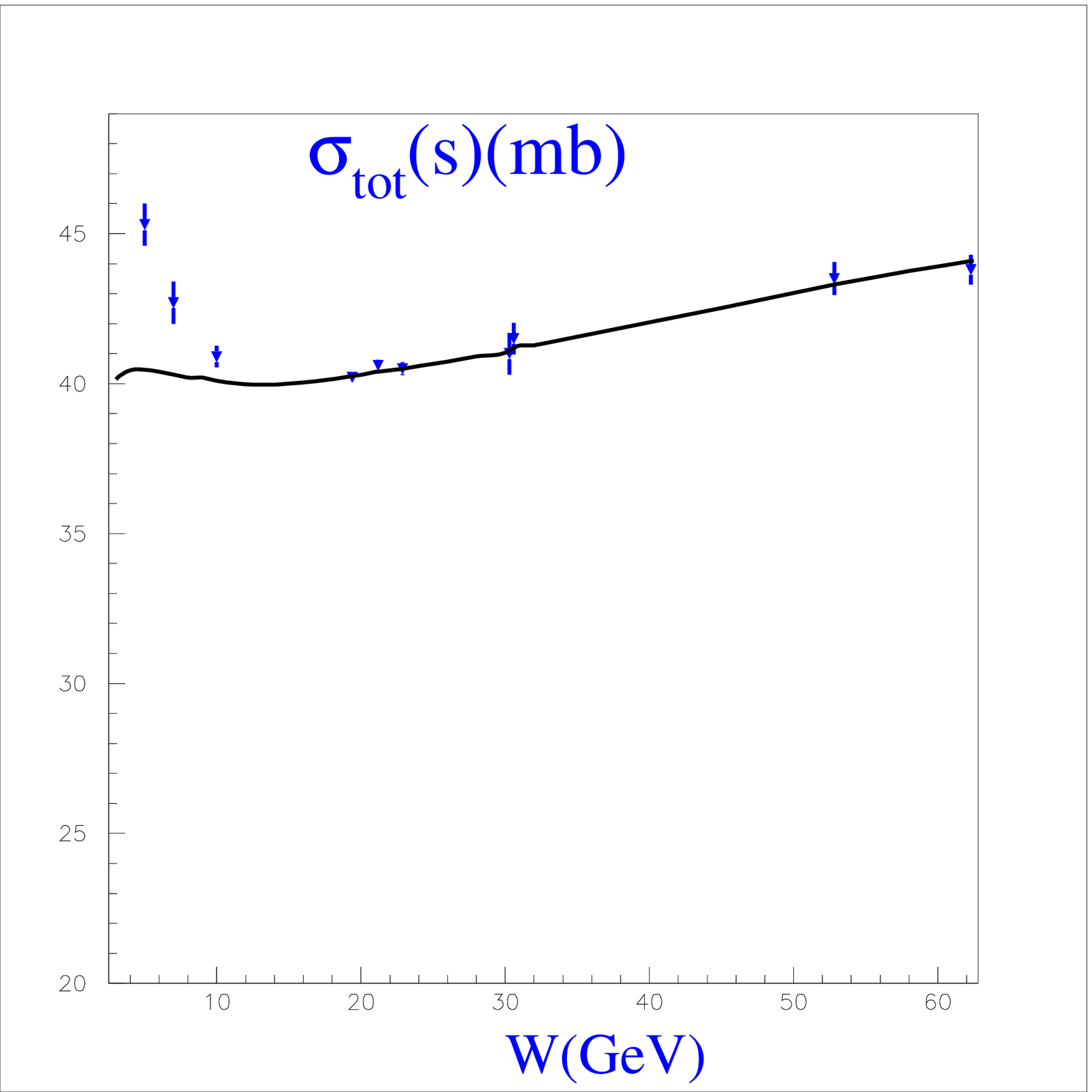,width=85mm}}
\caption{The total cross section ($\sigma_{tot} = 1/2(\sigma_{tot}(pp)
 + \sigma_{tot}(p \bar{p}))$) at low energies. The curve
 illustrates our parametrization. }
\label{lowe}}
\end{minipage}}

Our  fit is  based on 55 experimental data points,
which includes the $p$-$p$ and $\bar{p}$-$p$ total cross sections,
integrated elastic cross sections,
integrated single and double  diffraction cross sections,
and the forward slope of the elastic cross section
in the ISR-Tevatron energy range. The model gives
a good reproduction of the
data,  with a $\chi^2/d.o.f. \,\approx 1.25$.
A significant contribution
to $\chi^2/d.o.f. $ stems from the uncertainty for the value of two single
diffraction cross sections, and of the total cross section
at the Tevatron, without these three points we achieve a 
$\chi^2/d.o.f.$ = 1.0.
In addition to the quantities contained in the data base, 
we obtain a good description of the CDF \cite{CDF} 
differential elastic cross sections and the single diffractive mass 
distribution at $t\; =\; 0.05\;GeV^{2}$.  An important advantage
of our approach,  is that
the model provides a very good reproduction of the double diffractive (DD) 
data points. Other attempts 
to describe the DD data e.g.(see Refs.\cite{KMR5}),
were not successful in reproducing the DD
experimental results over the whole energy range.

The predicted energy behaviour of the above mentioned cross sections, 
as well as the forward slope of the elastic cross section, 
are given in detail in Ref.\cite{GLMM}. 
We mention here only our results for the LHC energy, which are
of particular interest i.e. 
at W=14 $TeV$ we have  $\sigma_{tot}= 92.1\,mb$,
$\sigma_{el}=20.9\,mb$, $\sigma_{sd}=11.8 \,mb$,
$\sigma_{dd}=6.08\,mb$ and $B_{el}=20.6 \,GeV^{-2}$.

We note that the lowest energy data point in our 
fit\cite{GLMM} to the total cross 
section was at W = 19.2 GeV. In Fig. 3, we show the results of our 
calculations in which we have  
extrapolated our model to lower energies. 
As can be seen, it provides a fair 
description of the data even at energies as low as W = 7 GeV, where we 
underestimate the experimental numbers just by about 10{$\%$}.
\section{Survival probability for diffractive production of the Higgs 
boson.}
In our approach, the survival probability for the
diffractive production of the Higgs boson, 
can be written as a product of two factors, 
\beq \label{SPF}
\SP \,\,\,
=\,\, \langle\mid S^2_{enh} \mid \rangle\Lb \mbox{enhanced diagrams}\Rb\,
\times\,\langle\mid S^2_{2ch} \mid \rangle\Lb \mbox{Good-Walker 
mechanism}\Rb.
\eeq

The second factor has been determined in a number of papers, and all
the  groups obtain approximately the same values.
The first factor is new and was
first evaluated in Ref.\cite{GLMM}, where the exact
formula for $\langle\mid S^2_{enh} \mid \rangle$ was derived. 
It has the form
\bea
&&\langle \mid S^2_{enh}\Lb MPSI\Rb \mid \rangle \left(Y\right)\,\,\,
=  \label{SPE51} \\
&&\,\,\,\,\,\,\,\,\,\,\,\,\,\,\,\,\,\,\,\,\,\,\,\,
= \,\,S\Lb {\cal T}(Y,Y')\Rb \,\,-\,\,2\,
e^{ - \Delta_\pom (Y - \delta Y_H)/2}\,S_{1}\Lb {\cal T}(Y,Y')\Rb\,\,+\,\,
e^{ - 2\Delta_\pom (Y - \delta Y_H)/2}\,S_{2}\Lb {\cal T}(Y,Y')\Rb; 
\notag\\
&& \,\,\,\,\,\,\,\,\,\,S(T)
= \frac{1}{T^3}\left\{ - T\,+\,e^{\frac{1}{T}}\,\Lb 1  + T \Rb \,
e^{\frac{1}{T}}\,\Gamma\Lb 0,  \frac{1}{T}\Rb\right\}; \notag\\
&& \,\,\,\,\,\,\,\,\,S_{1}(T) =
\frac{1}{T^3}\left\{   -T (1 + T) \,+\, ( 1 + 2T)\,
e^{\frac{1}{T}}\,\Gamma\Lb 0, \frac{1}{T}\Rb\right\};\label{SPE52}\\
&& \,\,\,\,\,\,\,\,\,S_{2}(T) =
\frac{1}{T^3} \left\{ T\left[ \Lb  T  - 1 \Rb^2 - 2 \right]  \,
+\, ( 1 + 3 T)\, e^{\frac{1}{T}}\,\Gamma\Lb 0,
\frac{1}{T}\Rb\right\},
\label{SPE53}
\eea
where, 
\beq \label{TT}
{\cal T} \Lb Y,Y'\Rb\,\,\,=\,\,\gamma \Lb 
e^{\Delta_\pom ( Y - Y')} \,-\, 1\Rb\,\Lb e^{ \Delta_\pom Y'} \,-\, 1\Rb.
\eeq
This factor diminishes the value of the survival probability,
and causes a decrease in its magnitude by a factor of 5-6, in going from 
Tevatron  to  LHC energies (see Table ~\ref{t3}).
\TABLE[ht]{
\begin{tabular}{||l|l|l|l||}
\hline \hline
  &  \,\,\,\,\,\,\,\,\,\,\,\,\,\,\,\,\,\,\,\,\,\,\,\,Tevatron & 
\,\,\,\,\,\,\,\,\,\,\,\,\,\,\,\,\,\,\,\,\,\,\,\, LHC &
\,\,\,\,\,\,\,\,\,\,\,\,\,\,\,\,\,\, $W= 10^5 GeV$  \\
 & GLMM\,\,\,\,\,\,\,\,\,\,\,\,\,\,\,\,\,\,\,\,\,\,\,\,KMR & 
GLMM\,\,\,\,\,\,\,\,\,\,\,\,\,\,\,\,\,\,\,\,\,\,\,\,KMR & 
GLMM\,\,\,\,\,\,\,\,\,\,\,\,\,\,\,\,\,\,\,\,\,\,\,\,KMR\\\hline
$\sigma_{tot}$( mb ) & 73.29 \,\,\,\,\,\,\,\,\,\,\,\,\,\,\,\,\,\,\,\,\,\,\,\,
74.0 
& 92.1\,\,\,\,\,\,\,\,\,\,\,\,\,\,\,\,\,\,\,\,\,\,\,\,\,\,88.0 & 
108.0\,\,\,\,\,\,\,\,\,\,\,\,\,\,\,\,\,\,\,\,\,\,\,\,98.0 \\ \hline
$\sigma_{el}$(mb) & 16.3 \,\,\,\,\,\,\,\,\,\,\,\,\,\,\,\,\,\,\,\,\,\,\,\,\,\,\,
16.3 & 
20.9 \,\,\,\,\,\,\,\,\,\,\,\,\,\,\,\,\,\,\,\,\,\,\,\,20.1 & 
24.\,\,\,\,\,\,\,\,\,\,\,\, \,\,\,\,\,\,\,\,\,\,\,\,\,\,\,22.9 \\\hline
$\sigma_{sd}$(mb) & 9.76 \,\,\,\,\,\,\,\,\,\,\,\,\,\,\,\,\,\,\,\,\,\,\,\,\,\,
10.9 & 
11.8 \,\,\,\,\,\,\,\,\,\,\,\,\,\,\,\,\,\,\,\,\,\,\,\,13.3 &
14.4  \,\,\,\,\,\,\,\,\,\,\,\,\,\,\,\,\,\,\,\,\,\,\,\,15.7 \\
\,\,\,\,\,\,\,\,\,\,\,\,\,\,\,\,\,\,\,\,\,\,\,\,\,\,\,\,\,\,\,\,
$\sigma^{\mbox{low M}}_{sd}$ & 
8.56\,\,\,\, \,\,\,\,\,\,\,\,\,\,\,\,\,\,\,\,\,\,\,\,\,\,\,4.4 & 
10.52\,\,\,\, \,\,\,\,\,\,\,\,\,\,\,\,\,\,\,\,\,\,
5.1 & 12.2\,\,\,\, \,\,\,\,\,\,\,\,\,\,\,\,\,\,\,\,\,\,\,\,\,\,\,\,5.7 \\
\,\,\,\,\,\,\,\,\,\,\,\,\,\,\,\,\,\,\,\,\,\,\,\,\,\,\,\,\,\,\,\,
$\sigma^{\mbox{high  M}}_{sd}$ 
& 1.2\,\,\,\, \,\,\,\,\,\,\,\,\,\,\,\,\,\,\,\,\,\,\,\,\,\,\,\,\,\,\,6.5 & 
1.28\,\,\,\, \,\,\,\,\,\,\,\,\,\,\,\,\,\,\,\,\,\,\,\,\,\,\,8.2 & 
2.2\,\,\,\,\,\,\,\,\,\,\,\,\,\,\,\, \,\,\,\,\,\,\,\,\,\,\,\,10.0 \\ \hline
$\sigma_{dd}$(mb) & 5.36
\,\,\,\,\,\,\,\,\,\,\,\,\,\,\,\,\,\,\,\,\,\,\,\,\,\,7.2 &  
6.08 \,\,\,\,\,\,\,\,\,\,\,\,\,\,\,\,\,\,\,\,\,\,\,\,13.4 &
6.29 \,\,\,\,\,\,\,\,\,\,\,\,\,\,\,\,\,\,\,\,\,\,\,\,17.3 \\
 \hline
$\Lb\sigma_{el} + \sigma_{sd} + \sigma_{dd}\Rb/\sigma_{tot}$ & 
0.428 \,\,\,\,\,\,\,\,\,\,\,\,\,\,\,\,\,\,\,\,\,\,\,\,0.464 & 
0.421 \,\,\,\,\,\,\,\,\,\,\,\,\,\,\,\,\,\,\,\,\,\,\,\,0.531 & 
0.412\,\,\,\,\,\,\,\,\,\,\,\,\,\,\,\,\,\,\,\,\,\,\,\,\,\,0.57 \\ \hline
$S^2_{2ch}(\%)$ & 
3.2 (2.7) \,\,\,\,\,\,\,\,\, \,\,\,\,\,\,\,\,\,\,\,\,\,\,2.7 - 4.8 & 
2.35 (0.45) \,\,\,\,\,\,\,\,\, \,\,\,\,\,1.2-3.2 &
2.0  \,\,\,\,\,\,\,\,\,\,\,\, \,\,\,\,\,\,\,\,\,\,\,\,\,\, 0.9 - 2.5 \\
\hline
$S^2_{enh}(\%)$ & 28.5 (100)\,\,\,\,\,\,\,\,\, \,\,\,\,\,\,\,\,\,\,\,\,100 & 
6.3  (100)\,\,\,\,\,\,\,\,\, \,\,\,\,\,\,\,\,\,\,\,100 & 
3.3 \,\,\,\,\,\,\,\,\,\,\,\, \,\,\,\,\,\,\,\,\,\,\,\,\,\,100 \\ \hline
$S^2(\%)$ & 0.91  (2.7) \,\,\,\,\,\,\,\,\,\,\,\,\,\,2.7 - 4.8 & 
0.15 (0.45)\,\,\,\,\,\,\,\,1.2-3.2 &
0.066\, \,\,\,\,\,\,\,\,\,\,\,\, \,\,\,\,\,\,\,\,\,\,\,\,0.9 - 2.5\\
\hline \hline
\end{tabular}
\caption{Comparison of the GLMM (this paper) and KMR\cite{KMRNEW}
models. The numbers in parenthesis are the estimates
for the contribution of the semi-enhanced
diagrams to the survival probability.}
\label{t3}}

After hearing our report  on  the above results at the "Diffraction'08" 
conference, Alexei Kaidalov posed two questions,
(also see his talk at International Symposium on Multiparticle Dynamics 2008).  
The first concerned the so called, threshold effect, which suggests 
that one can only use the one Pomeron exchange approximation at sufficiently
large rapidities ($Y > 2$). In our approach there is no need for an additional
cutoff in rapidity, since  the scattering
amplitude is described by  the sum of the contributions of both  Pomerons 
and Reggeons.

In \fig{lowe} we show the scattering  cross section at rather low energies. 
Our parametrization provides a good description  
of the experimental data also at  low energies.
Since in our parametrization we include the contribution of the secondary 
Reggeons, there is no need for  an additional cutoff.
The second question concerns the semi-enhanced diagrams
of the \fig{enh}-b type. Indeed, in our fit since $g_1$ and $g_2$ turn out
to be large, we could anticipate a large contribution.  
A self consistent approach which includes  all Pomeron diagrams will
be published in a separate paper, which is presently 
being prepared\cite{GLMM2}. 
In the present paper we only calculate the exclusive contribution of the 
diagrams of type \fig{enh}-b, given by the following formula
(the proof of this formula will be given in Ref.\cite{GLMM2})
\bea
 \SP\,\,&=&\,\,\frac{1}{\int d^2 b \,a^2_H(b)}\,\int\,d^2 b\,\,
 \left\{\int d^2 b' \,\frac{\,a_H^2(b)}
{\Lb 1\,\,+\,\,g_i\,\Gamma(b';m_i)\,T(\gamma, \Delta,y)\,\,+\,\,\,g_k\,
\Gamma(|\vec{b} - \vec{b}'|;m_i)\,T(\gamma, \Delta,y)\Rb^2}\right\}  
\nonumber\\
 &\times &\Lb 1 -\exp\Lb  \int d^2 b' \,\frac{g_i\,g_k \,\Gamma(b';m_i)\,
\Gamma(|\vec{b} - \vec{b}'|;m_i)\,T(\gamma,
 \Delta,y)}{1\,\,+\,\,g_i\,\Gamma(b';m_i)\,T(\gamma, \Delta,y)\,\,
+\,\,\,g_k\,\Gamma(|\vec{b} - \vec{b}'|;m_i)\,T(\gamma, \Delta,y)}\Rb\Rb.  
\label{SPFS}
\eea
In \eq{SPFS} $T(\gamma,\Delta,y)$ is given by \eq{T}, and 
\beq \label{G}
\Gamma\Lb b, m_i\Rb =\,\frac{m^3_i\,b}{4 \pi}\,K_1\Lb m_i,b\Rb
\eeq
where $K_1 $ is the modified Bessel function of the second kind.
The vertices $g_i$, masses $m_i$ as well as the values of
the parameters $\gamma$ and $\Delta$ and the form
of the hard amplitude $A_H$ were taken from Ref. \cite{GLMM}.

The numerical estimates, in which the semi-enhanced contribution  
(Eqn. (4.6)) was exclusively included, 
are given in parenthesis in Table~\ref{t3}.
 The
semi-enhanced diagrams were neglected in \cite{GLMM}.
Our present calculation, with only the semi-enhanced diagrams, 
serves only as a reminder that our calculation in Ref.\cite{GLMM} is
incomplete. We expect, though, $S^{2}(semi-enhanced)$ 
to considerably reduce 
$\langle\mid S^2_{2ch} \mid \rangle\Lb \mbox{Good-Walker 
mechanism}\Rb$, especially at the LHC energy. Estimates which include both
semi-enhanced and enhanced diagrams will be published soon\cite{GLMM2}.

\FIGURE[h]{\begin{minipage}{90mm}
{\centerline{\epsfig{file=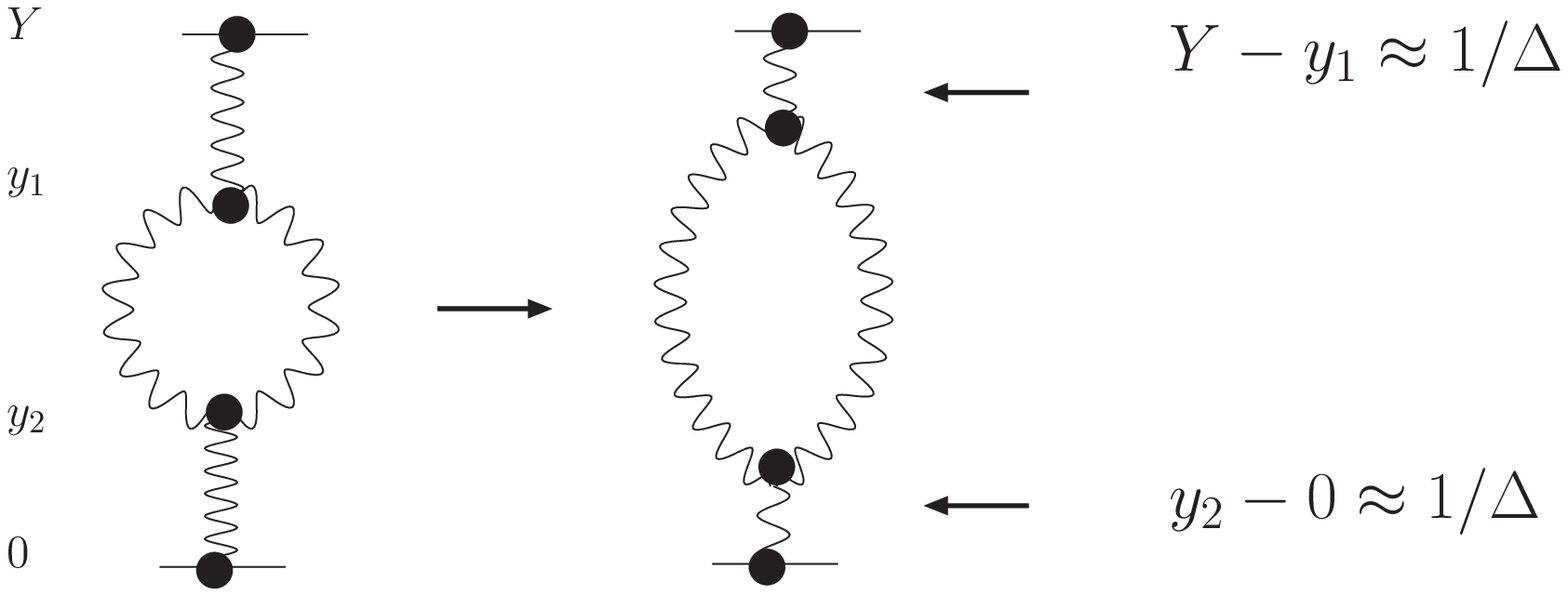,width=85mm}}
\caption{The first enhanced diagram.}
\label{enhd2}}
\end{minipage}}

In spite of the fact that we do not need to take 
the threshold effect into account, we estimate its influence on
the value of the survival probability. 
We assume that we can use  Pomeron exchange only for $Y >y_0$,
in this case the first enhanced diagram shown
in \fig{enhd2} has the following contribution
(see a detailed discussion in Refs.\cite{LEPR,LMP})
\beq \label{ENH2}
A\Lb \mbox{\fig{enhd2}}\Rb\,\,=\,\,\frac{g^2 \gamma^2}
{\Delta^2}e^{ - 2\,\Delta \Lb  Y - y_0\Rb}\,\,+\,
\,{\cal O}\Lb \gamma\, e^{ - \Delta Y} \Rb. 
\eeq 
Therefore, in order to introduce the threshold correction we need
to change $\gamma \to \gamma \exp \Lb \Delta y_0\Rb$,
and multiply the Pomeron contribution by the step function
$\Theta\Lb Y - y_0\Rb$. The first observation
means that in order to maintain 
the quality of the experimental data description  
obtained without the threshold effect,
we need to increase the value of $\gamma$ in
our parametrization by a factor $\exp\Lb \Delta y_0\Rb$.
We, then, need to take into account $\Theta\Lb Y - y_0\Rb$.
Doing so we obtain the following values for
$\langle\mid S^2_{enh} \mid \rangle$
at $y_0 = 0, 1.5, 2.3$: 
for the Tevatron energy we obtain values of 0.285, 0.7, 0.99
respectively, 
and for the LHC energy the corresponding values are 0.06, 0.12, 0.19
respectively. 
Our conclusion is that the threshold effect is not needed
in our approach. Regardless, 
its inclusion does not change 
the fact that the value of
$\langle\mid S^2_{enh} \mid \rangle$ 
decreases by a factor of 5-6  in going from the Tevatron to the LHC 
energy.
It should be stressed that since
$Y - y_1 \approx y_2 - 0 \approx 1/\Delta \approx 3$,  
the theoretical accuracy associated with 
this calculation is low.
\section{Conclusions}
In this letter we present a consistent theoretical approach that 
(i) provides a natural bridge both with a N=4 SYM
Pomeron with $\alpha'_\pom =0$, and with the pQCD approach; 
(ii) gives an analytical expression for summing the Pomeron interactions;
(iii) leads to a very good ($\chi^2/d.o.f \,= \, 1$)
description of all available data on strong interactions
at high energy and (iv) predicts survival probability to be less that 1\% 
at the LHC.

Within the framework of our approach, we still need to solve the
problems of 1) accounting for the semi-enhanced diagrams summed 
together with the enhanced ones (see Ref. \cite{GLMM2});
2) to develop a
theoretical approach for the Pomeron-hadron interactions; 
and 3) to formulate the Pomeron-Reggeon interaction on a more 
sound theoretical basis.

To illustrate our achievements and problems,
we compare our approach with the work of the Durham group (KMR), 
whose results were presented in six papers 
recently published\cite{KMRNEW,KMR5}.
Both approaches consider $\alpha'_\pom$  as being small.

In both programs  the Pomeron interaction was taken into account.  
The difference between the two approaches is that KMR 
made an {\it ad hoc} "reasonable" assumption, that
the  multi-Pomeron vertices have the following form,
for the transition of  $n$ Pomerons to $m$ Pomerons
\beq \label{DG1}
g^n_m\,=\,n\,m\,\lambda^{n + m -2}\,g_N/2\,
=\,n\,m\,\lambda^{n+m-3}\,g_{3{\pom}}/2.
\eeq
No theoretical arguments or theoretical models were offered in support
of this assumption, which certainly contradicts
the pQCD approach \cite{BART,BRN}.
We view this assumption as problematic,
since it eliminates the hope of building an
approach that matches with pQCD.
In paper 4 of Ref.\cite{KMR5}, the system of two equations
(Eq.(26) and Eq.(27) of this paper) has been suggested
for the opacity $\Omega^\pom$,
and it is stated that they follow from the parton model.
Since there is no proof of this claim, and while we are
inclined to believe the authors, that the equations could indeed
be proved, we should still bear in mind, that the parton approach
(see Refs.\cite{FEYN,GRIB} is based on the assumption
that there is only a short range rapidity interaction
between partons, while we, due to exchange of gluons in QCD,
have a long range rapidity interaction.

KMR numerically solved Eq.(5.1) using an iteration method.
The accuracy of the solution and its stability were not discussed.
Since our approach is a particular case of the KMR approach with only the
triple Pomeron interaction, we are surprised that the
interaction method works, since our Pomeron Green's
function $G^\pom \,\propto \sum_n (-1)^n\, n!\Lb  \gamma \exp(\Delta Y)\Rb^n$.
Such a series cannot not be reproduced by an iterative
procedure. Unfortunately, KMR do not provide any details of their 
numerical procedure.

For single diffraction in the region of high mass, KMR 
encounter the same problem as we have in this paper, since
they do not include the full set of semi-enhanced diagrams
(an example of a neglected diagram is the second diagram of \fig{enh}-b).

Although we consider the KMR approach to be much less 
theoretically reliable than ours, the numerical results regarding the 
elastic and total cross sections are very close. 
We consider this fact as confirmation
that most properties of the high energy interaction are related
to the Pomeron interaction, but not to the details of the interaction.

We disagree with KMR, regarding the predicted value   
of the survival
probability for central exclusive diffractive Higgs production.
In Ref.\cite{KMRNEW} the effect of enhanced and semi-enhanced
diagrams have been neglected, with some supportive arguments of
why it is reasonable approximation, however, in the last five papers
\cite{KMR5} some of the diagrams have been included. 
The main difference is that KMR consider the
threshold effect as being very important, and
by choosing  $Y > 2.3$, they  obtain a suppression factor 
0.83 for the Tevatron energy, and 0.67 for the LHC.

We have explained why in our approach, it is not necessary to
include the threshold effect separately. We 
have shown that the threshold effect does not change
the fact, that the value of the survival probability
decreases by a  larger factor in going from the Tevatron to the LHC 
energy. 
We are of the opinion that, KMR should, first, 
systematically  introduce the Reggeon contributions, and
only after that discuss the
threshold effect, both in the calculation of the  survival probability 
and in the scattering amplitude, especially for the case of single
diffraction production.
\section* {Acknowledgements}

This research was supported
in part by the Israel Science Foundation, founded
by the Israeli Academy of Science
and Humanities, by BSF grant $\#$ 20004019 and by
a grant from Israel Ministry of Science, Culture and Sport and
the Foundation for Basic Research of the Russian Federation.


\end{document}